\documentclass[preprint,
showpacs,
bibnotes,
amsmath,
amssymb,
aps,
prb,
floatfix ]{revtex4-1}

\usepackage{amsmath}
\usepackage{graphicx}
\usepackage{epstopdf}

\begin{document}

\title{Magnetic Switching Dynamics due to Ultrafast Exchange Scattering: A Model Study}

\author{Alexander Baral}
\affiliation{Physics Department and Research Center OPTIMAS, University of Kaiserslautern, 67663 Kaiserslautern, Germany}
\author{Hans Christian Schneider}
\email{hcsch@physik.uni-kl.de} 
\affiliation{Physics Department and Research Center OPTIMAS, University of Kaiserslautern, 67663 Kaiserslautern, Germany}

\date{\today}

\begin{abstract}
We study the heat-induced magnetization dynamics in a toy model of a ferrimagnetic alloy, which includes  localized spins antiferromagnetically coupled to an itinerant carrier system with a Stoner gap. We determine the one-particle spin-density matrix including exchange scattering between localized and itinerant bands as well as scattering with phonons. While a transient ferromagnetic-like state can always be achieved by a sufficiently strong excitation, this transient ferromagnetic-like state only leads to magnetization switching for model parameters that also yield a compensation point in the equilibrium $M(T)$ curve. 
\end{abstract}

% 72.25.Rb - Spin-Relaxation and Scattering
% 75.78.-n - Magnetization Dynamics
% 75.78.Jp - Ultrafast magnetization dynamics and switching
% 77.80.Fm - Switching phenomena 

\maketitle

\section{Introduction}

Heat-induced reversal of magnetization in ferrimagnetic alloys~\cite{Radu.2011,Ostler.2012} and multilayers~\cite{Bergeard.2014} is a recent development that has generated a lot of attention because it realizes an ultrafast \emph{deterministic} switching without magnetic fields that may pave the way to a faster magnetic logic. After the first phenomenological analysis of this switching process using the Baryakhtar equation,~\cite{Mentink.2012} a more microscopic understanding of this effect is now being developed using models that in one way or another include \emph{classical models of localized} spins on different sub-lattices, which are coupled by an exchange interactions between the different lattices \cite{Wienholdt.2013,Atxitia.2013,Schellekens.2013} The atomistic models usually involve a thermal bath averaging, which is also implicit in Landau-Lifshitz-Bloch type calculations.~\cite{Garanin.1997,Nowak.2007, Kazantseva.2008} Ref.~\onlinecite{Schellekens.2013} does include \emph{itinerant} carrier-phonon scattering, but this model is based on a non-standard electron-phonon Hamiltonian and separates the charge degrees of electrons from their spins so that its two spin systems essentially are also two types of localized spins coupled to a bath. Although existing theoretical approaches have established the importance of an exchange coupling between localized magnetic moments on different sublattices for heat-induced magnetic switching and the occurrence of transient ferromagnetic-like states, the microscopic picture of magnetization switching is still unclear, as the phenomenological model~\cite{Mentink.2012} implies a counterintuitive interplay between spin-orbit (``relativistic'') and exchange scattering. Further, there are differences between existing atomistic calculations, some of which stress~\cite{Atxitia.2013} that \emph{transverse} magnetization dynamics occurs during switching, while others do not find such a transverse magnetization component after bath averaging.~\cite{Wienholdt.2013} 

\section{Model and equilibrium configuration}

In this paper we put forward a different microscopic model for magnetization dynamics---inspired by theories~\cite{Jungwirth.2006,Cywinski.2007} for magnetic semiconductors---that (i) includes an exchange interaction between delocalized and localized electrons \emph{in a band picture} and (ii) that is capable of including the coupling of the carriers to the environment, i.e., phonons, in a microscopic fashion. The model used in the following contains two bands of different carrier species which have different spin and are designed to resemble itinerant $3d$ electrons in iron and localized $4f$ electrons in gadolinium, respectively. For simplicity, we assume a spin of $s=1/2$ and a parabolic dispersion for the itinerant carriers, as well as a spin $S=1$ and a completely flat band for the localized states; we thus ignore complications that arise from a finite width of the bands which is present in real materials. Denoting the localized spin states by $|\nu\rangle=\pm1,0$ and the electron states by $|\vec{k} \sigma\rangle$, where $\sigma=\pm$, we calculate dynamically the spin-resolved one-particle density matrices of the itinerant electrons $\rho^{\sigma \sigma'}_{k}$ and the localized spins $\rho^{\nu \nu'}_{\text{loc}}$ with an equation of motion technique. Using the spin-dependent reduced density matrices we do not separate the charge and the spin of the itinerant electrons, as done in models that work with three temperatures, most notably with different spin and electron temperatures. ``Magnetic'' contributions are an antiferromagnetic exchange interaction $J$ between itinerant and flat bands and a Stoner-like on-site coupling $U$ among itinerant carriers. The latter is treated in mean-field approximation and favors a ferromagnetic electron spin polarization. Long and short range ($U$) contributions to electronic Coulomb scattering are neglected because (i) they do not change the itinerant spin polarization, and (ii) for the conditions studied here, the electronic distributions never deviate much from Fermi-Dirac distributions.~\cite{Mueller.2013} By contrast, we include both mean field and scattering contributions from the exchange interaction between both localized and itinerant spins, as well as a coupling of the itinerant carriers to acoustic phonons at the level of Boltzmann scattering integrals. Throughout, we work with single-particle states that are obtained from a diagonalization including the exchange and Stoner mean-field contributions. The resulting mean-field energies are denoted by $E_{\nu}$ for the localized spins and $\epsilon_{\vec{k} \sigma}$ for the spin-split electron bands. The relevant equations of motion then take the form 
\begin{align}
		\frac{\partial}{\partial t} \rho_{\text{loc}}^{\nu \nu'} &= \frac{i}{\hbar} (E_{\nu} - E_{\nu'}) \rho_{\text{loc}}^{\nu \nu'} + \frac{\partial}{\partial t} \rho_{\text{loc}}^{\nu \nu'}\big|_{\text{xc}}	\label{eq:EOM_Loc}\\
\begin{split}\frac{\partial}{\partial t} \rho_{\vec{k}}^{\sigma \sigma'} & = \frac{i}{\hbar} (\epsilon_{\vec{k} \sigma} - \epsilon_{\vec{k} \sigma}) \rho_{\vec{k}}^{\sigma \sigma'}
	+ \frac{\partial}{\partial t} \rho_{\vec{k}}^{\sigma \sigma'}\big|_{\text{xc}}
 	\quad + \frac{\partial}{\partial t} \rho_{\vec{k}}^{\sigma \sigma'}\big|_{\text{e-pn}}
   	- \frac{\rho_{\vec{k}}^{\sigma \sigma'} - \delta_{\sigma \sigma'} F_{\vec{k}}^{\sigma}}{\tau_{\text{sf}}}
   	\end{split}
	\label{eq:EOM_Itin}
\end{align}
The first terms in Eq.~\eqref{eq:EOM_Loc} and \eqref{eq:EOM_Itin} are coherent contributions, which describe precessional motion of one spin around the mean-field of the spin of the other species. All the other terms are incoherent terms. The incoherent exchange scattering contributions at the level of Boltzmann scattering integrals are given by
\begin{widetext}
\begin{equation}
	\label{Loc_ExchangeBoltzmann}
	\begin{aligned}
	\left.\frac{\partial}{\partial t} \rho_{\text{loc}}^{\nu_1 \nu_2}\right|_{\text{xc}} =& 
	\frac{i}{\hbar} \sum_{\vec{k} \vec{k}'} \sum_{\sigma_{1} \ldots \sigma_{4}} \sum_{\nu_{3} \nu_{4} \nu_{5}} W^{\nu_{5} \nu_{1}}_{\vec{k}' \sigma_{1} \vec{k} \sigma_{2}} (W^{\nu_{3} \nu_{4}}_{\vec{k}' \sigma_{3} \vec{k} \sigma_{4}})^{*} \\ & \cdot \frac{ \delta_{\nu_4 \nu_2} \rho_{\text{loc}}^{\nu_{5} \nu_{3}} \rho_{\vec{k}'}^{\sigma_{1} \sigma_{3}} (\delta_{\sigma_{4} \sigma_{2}} - \rho_{\vec{k}}^{\sigma_{4} \sigma_{2}}) - \delta_{\nu_{5} \nu_3} \rho_{\text{loc}}^{\nu_{4} \nu_{2}} \rho_{\vec{k}}^{\sigma_{4} \sigma_{2}} (\delta_{\sigma_{1} \sigma_{3}} - \rho_{\vec{k}'}^{\sigma_{1} \sigma_{3}}) } { \epsilon_{\vec{k} \sigma_{4}} - \epsilon_{\vec{k}' \sigma_{3}} + E_{\nu_{4}} - E_{\nu_{3}} - i \gamma } + ( \nu_1 \leftrightarrow \nu_2)^{*}
	\end{aligned}
\end{equation}
\begin{equation}
	\label{Itin_ExchangeBoltzmann}
	\begin{aligned}
	\left.\frac{\partial}{\partial t} \rho_{\vec{k}}^{\sigma_1 \sigma_2}\right|_{\text{xc}} =& 
	\frac{i}{\hbar} \sum_{\vec{k}'} \sum_{\nu_{1} \ldots \nu_{4}} \sum_{\sigma_{3} \sigma_{4} \sigma_{5}} W^{\nu_{1} \nu_{2}}_{\vec{k}' \sigma_{5} \vec{k} \sigma_{1}} (W^{\nu_{3} \nu_{4}}_{\vec{k}' \sigma_{3} \vec{k} \sigma_{4}})^{*} \\ & \cdot \frac{\delta_{\nu_4 \nu_2} \rho_{\text{loc}}^{\nu_{1} \nu_{3}} \rho_{\vec{k}'}^{\sigma_{5} \sigma_{3}} (\delta_{\sigma_{4} \sigma_{2}} - \rho_{\vec{k}}^{\sigma_{4} \sigma_{2}}) - \delta_{\nu_1 \nu_3} \rho_{\text{loc}}^{\nu_{4} \nu_{2}} \rho_{\vec{k}}^{\sigma_{4} \sigma_{2}} (\delta_{\sigma_{3} \sigma_{5}} - \rho_{\vec{k}'}^{\sigma_{3} \sigma_{5}})} {\epsilon_{\vec{k} \sigma_{4}} - \epsilon_{\vec{k}' \sigma_{3}} + E_{\nu_{4}} - E_{\nu_{3}} - i \gamma} + ( \sigma_1 \leftrightarrow \sigma_2)^{*}
	\end{aligned}
\end{equation}
\end{widetext}
Here, $\gamma$ denotes an infinitesimal broadening and we use the abbreviation $W^{\nu \nu'}_{\vec{k} \sigma \vec{k}' \sigma'} \equiv J \; \langle \vec{k} \sigma | \hat{\vec{s}} | \vec{k}' \sigma' \rangle \langle \nu | \hat{\vec{S}} | \nu' \rangle$, where~$\hat{\vec{s}}$ and $\hat{\vec{S}}$ are the spin-operators of the localized and the itinerant electrons. At this point, a comment regarding the exchange parameter $J$ may be in order. In our calculation, $J$ is a Coulomb matrix element that occurs in the hamiltonian and directly enters the equation of motion for quantum-mechanical correlation functions; it could be calculated directly from the true electronic Bloch or Wannier functions. In atomistic spin models the $J$s are the parameters of an \emph{effective classical Heisenberg model} that are extracted from ab-initio electronic structure calculations by computing the (exchange) energy changes for a small tilting of the magnetic moments in adjacent unit cells.~\cite{Liechtenstein.1987} 

The electron-phonon scattering (e-pn) contribution in Eq.~\eqref{eq:EOM_Itin} is a standard expression, an explicit derivation of which with special attention to spin splitting is given in Ref.~\onlinecite{Baral.2014}. It is an important property of scattering with long-wavelength longitudinal phonons that it does \emph{not} lead to a transfer of angular momentum from the itinerant carriers to the lattice, it only cools down the itinerant carrier system and increases the temperature of the phonon system in accordance with energy conservation. The transfer of angular momentum to the lattice is left to a relaxation-time expression in Eq.~\eqref{eq:EOM_Itin} because the fundamental mechanism is not the important point of this paper. Likely it is a combination of electron-phonon/electron-electron scattering and spin-orbit coupling.~\cite{Krauss.2009, Mueller.2013} We generally assume spin-flip processes to be faster than the heat transfer to the phonons, but slower than the exchange scattering. In Eq.~\eqref{Itin_ExchangeBoltzmann}, $F_{\vec{k}}^{\sigma}=f(\epsilon_{\vec{k} \sigma}-\mu_F)$ denotes a Fermi-Dirac distribution with the same energy as the actual~$\rho_{\vec{k}}^{\sigma \sigma'}$, but equal chemical potentials $\mu_F$ for both spins. 

We determine the equilibrium configuration self-consistently, assuming that the equilibrium reduced density matrix~$\rho_{\vec{k}}^{\sigma \sigma'}=\delta_{\sigma,\sigma'}f(\epsilon_{\sigma\vec{k}}-\mu_0)$ is spin diagonal and given by Fermi-Dirac distributions with temperature $T_0$  and equal chemical potentials $\mu_0$ for both spin states.  The average spins $\vec{S}$ and $\vec{s}$ of the two species are parallel to the mutual exchange field. Depending on the value of the coupling constants $J$ and $U$ the itinerant system is either partially or fully spin polarized. In the numerical calculations below, as done in Ref.~\onlinecite{Baral.2014}, we employ two-dimensional $\vec{k}$ vectors because it reduces the numerical complexity of propagating the scattering calculations over long times. This simplification also changes the equilibrium spin polarization, exchange splitting and the Curie temperature compared to three-dimensional $\vec{k}$ vectors. In the following numerical calculations we always assume a common initial temperature of all three sub-system $T_{0} = 10\,\text{K}$, which is far lower than the Curie temperature $T_\text{C}$. 

\section{Results for magnetization dynamics}

We model the excitation of a short linearly-polarized laser pulse as an instantaneous heating of the itinerant carriers at $t=0$. We assume that the localized spin system is not excited optically.~\cite{Manchon.2012} Immediately after the excitation the itinerant spin density-matrix $\rho_{\vec{k}}^{\sigma \sigma'}$ is assumed to be spin diagonal with the spin dependent distributions determined by Fermi functions with the same spin-expectation value but with an elevated initial temperature~$T_\text{e}^{(0)}$ that usually exceeds $T_\text{C}$. Even though the spin polarization of the carriers does not change, the chemical potentials $\mu_\sigma$ become different.

%%%%%%%%%%%%%%%%%%%%%%%%%%%%%%%%%%%%%%%%%%%%%%%%%%
% RESULTS %
%%%%%%%%%%%%%%%%%%%%%%%%%%%%%%%%%%%%%%%%%%%%%%%%%%

\begin{figure}[tb]
	\centering
		\includegraphics[width=0.7\textwidth]{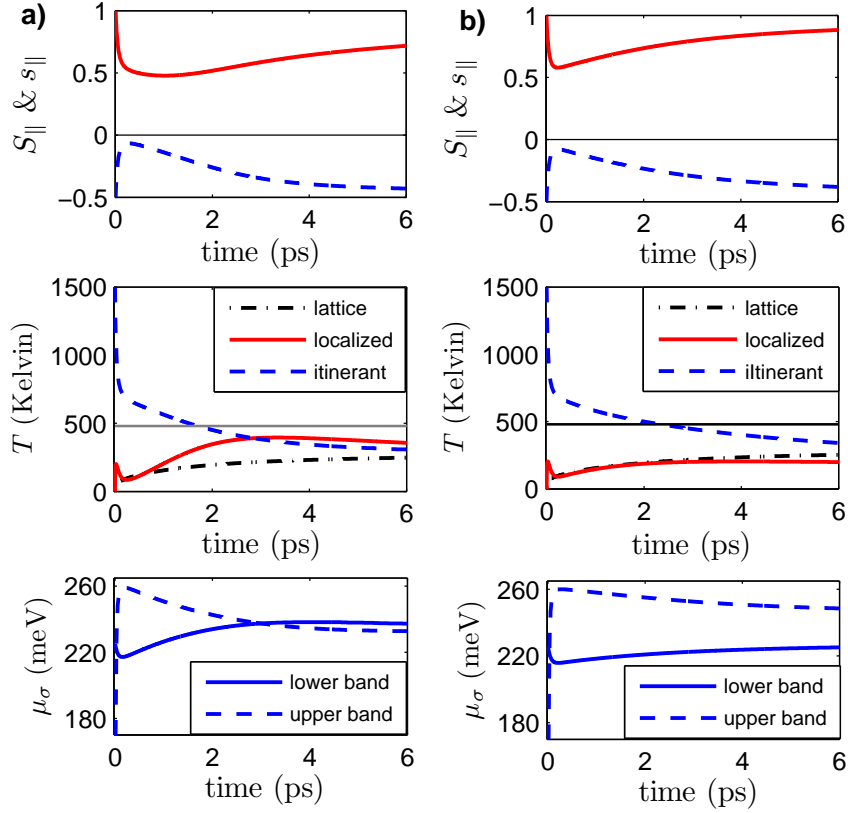}
	\caption{Computed results for demagnetization scenario with $J=100$\,meV, $U = 400\;\text{meV}$, and  $T_{\text{e}}^{(0)} = 1500\,\text{K}$. The spin-flip times are $\tau_{\text{sf}} = 200\,\text{fs}$ (a) and $\tau_{\text{sf}} \rightarrow \infty$ (b). Top: Dynamics of the averaged localized spin $S_{\parallel}$ (solid curve) and itinerant spin $s_{\parallel}$ (dashed curve); middle: quasi-equilibrium temperatures and $T_{\text{C}}$ (thin line); bottom: quasi-chemical potentials for the two itinerant bands.}
	\label{fig:3TempDemag}
\end{figure}

Figure~\ref{fig:3TempDemag} presents the key dynamical quantities for an excitation characterized by $T_{\text{e}}^{(0)} = 1500\,\text{K}$ as well as a short and infinitely long spin-flip time, respectively. The parameters  $J=100$\,meV and $U=400$\,meV give rise to a Curie temperature of $T_{\text{C}}=480$\,K. These parameters together with the initial carrier temperature $T_{\text{e}}^{(0)}$ lead to a demagnetization scenario, regardless of the spin-flip time $\tau_{\text{sf}}$, as shown in the top panel. The components of both localized and itinerant spins parallel to the mutual exchange field, $S_{\parallel}$ and $s_{\parallel}$, show an \emph{ultrafast} symmetric decrease due to exchange scattering on a timescale of several ten femtoseconds. The exchange scattering conserves the total angular momentum and energy of the combined system of spin-split itinerant carriers and localized spins, so that such an ultrafast drop does not occur for a ferromagnetic exchange coupling.~\cite{Mentink.2012} We have checked this also for our model. We will analyze the ultrafast dynamics due to the exchange scattering in more detail below.

Compared to the intrinsic time scale of the exchange scattering, spin-flip scattering, which dissipates only carrier angular momentum, and carrier-phonon scattering, which only transfers heat from the itinerant carriers to the phonon system, act on much longer time scales of hundreds of fs and several ps, respectively. During the comparatively slow remagnetization process, the angular momentum and energy transfer between the localized and itinerant sub-systems due to the exchange scattering is limited to the times scales of these slower mechanisms. Changing the spin-flip scattering time in Fig.~\ref{fig:3TempDemag}(b) as compared to (a) does not change the qualitative behavior. 

In the middle panel of Fig.~\ref{fig:3TempDemag} we show the quasi-equilibrium temperatures of the itinerant carriers (or ``electrons'') $T_{\text{e}}$, the localized spins $T_{\text{loc}}$ and the phonons $T_{\text{L}}$, which are obtained from the computed dynamics of the spin density matrix as the temperatures of thermalized distributions with the same energy as the non-equilibrium distributions. Note, in particular, that the excited electrons have a quasi-equilibrium temperature $T_{\text{e}}(t)$ and \emph{different} chemical potentials $\mu_\sigma (t)$ for each spin species, as shown in the bottom panel of Fig.~\ref{fig:3TempDemag}. With a non-vanishing spin-flip rate $\tau^{-1}_{\text{sf}}$ as in Fig.~\ref{fig:3TempDemag}(a), the temperatures of the itinerant electrons $T_{\text{e}}$ and of the localized spins $T_{\text{loc}}$ essentially converge on the time scale of the spin-flip relaxation $\tau_{\text{sf}}$. The phonon temperature $T_{\text{L}}$ approaches these two on the time scale of the electron-phonon scattering. Figure~\ref{fig:3TempDemag}(b) shows that without dissipation of angular momentum, viz.~$\tau_{\text{sf}} \rightarrow \infty$, $T_\text{loc}$ does not get close to $T_\text{e}$. This makes clear that exchange scattering neither simply equalizes the temperatures $T_\text{loc}$ and $T_\text{e}$, nor the chemical potentials $\mu_\sigma$. 

\begin{figure}[tb]
	\centering
		\includegraphics[width=0.7\textwidth]{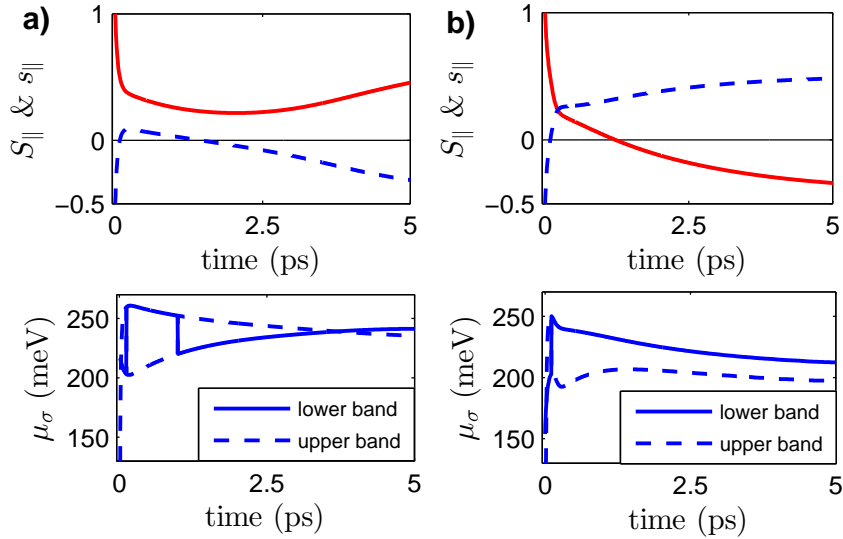}
	\caption{Top: Computed dynamics of the average localized spin $S_{\parallel}$ (solid curve) and the average itinerant spin $s_{\parallel}$ (dashed curve) for a TFS scenario with $T_{\text{e}}^{(0)} = 2000\,\text{K}$ and $U= 400\;\text{meV}$ (a) as well as a switching scenario with $T_{\text{e}}^{(0)} = 2500\,\text{K}$ and $U=500\;\text{meV}$ (b). Bottom: Electronic quasi-chemical potentials. $J=100$\,meV is unchanged.}
	\label{fig:TFS_SW}
\end{figure}

We next model stronger excitations by raising $T_{\text{e}}^{(0)}$ in Fig.~\ref{fig:TFS_SW}. The parameter $U$ in Fig.~\ref{fig:TFS_SW} (a) is the same as before so that we have $T_\text{C} \simeq 480$\,K. In addition, we relax the lattice temperature toward $T_0$ with a time constant of 10\,ps to include the effect of heat diffusion. This guarantees a final remagnetization without affecting the faster dynamics. For Fig.~\ref{fig:TFS_SW}(b) we increase $U$ to 500\,meV, which leads to $T_\text{C} \simeq 1200$\,K. For the same material parameters as in Fig.~\ref{fig:3TempDemag}(a), Fig.~\ref{fig:TFS_SW} exhibits a TFS starting around 50\,fs and persisting up to about 2\,ps.  During the TFS, the localized spins experience a population inversion and the corresponding spin-temperature $T_\text{loc}$ is no longer well defined. As the coupling to the phonons cools down the itinerant carriers, the system either returns, as in Fig.~\ref{fig:TFS_SW}(a), into its initial state or, as in Fig.~\ref{fig:TFS_SW}(b), remagnetizes with an inverse orientation and thus undergoes magnetization switching (SW). Note that in Fig.~\ref{fig:TFS_SW}(b) the itinerant carriers become fully spin-polarized, $s_\| \simeq 1/2$, around 2.5\,ps so that the exchange scattering cannot further remagnetize the localized spins. We label the eigenstates according to their energy, so that this label changes if the effective mean-field splitting changes its sign. Comparing the bottom panels of Figs.~\ref{fig:3TempDemag} and~\ref{fig:TFS_SW} indicates that the difference between the chemical potentials does not drive the switching or demagnetization dynamics.

\begin{figure}[tb]
	\centering
	\includegraphics[width=0.7\textwidth]{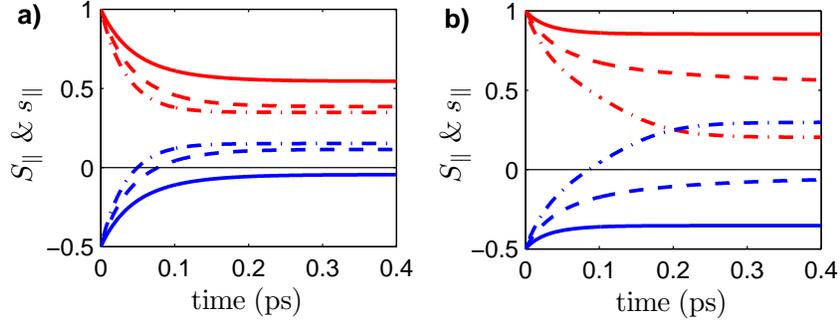}
	\caption{Computed dynamics of average localized spin $S_\|$ (red, starting from $1$) and average itinerant spin (blue, starting from $-0.5$) after excitation with $T_{\text{e}}^{(0)} = 1500$\,K (solid lines), $T_{\text{e}}^{(0)} = 2000$\,K (dashed lines) and $T_{\text{e}}^{(0)} = 2500$\,K (dashed-dotted lines) for $U=400$\,meV (a), $U=500$\,meV (b), and $J=100$\,meV.}
	\label{fig:Exchange_only}
\end{figure}

In Fig.~\ref{fig:Exchange_only} we come back to the role of the exchange scattering in realizing the TFS. We plot here for different initial temperatures $T_{\text{e}}^{(0)}$ the computed dynamics obtained by including only exchange scattering. While this leads to an unphysical steady state, it illustrates the way in which exchange scattering works. In Fig.~\ref{fig:Exchange_only}(a), for $U=400$\,meV,  an initial itinerant temperature $T_{\text{e}}^{(0)}=1500$\,K leads to a state with reduced localized spin, i.e., demagnetization in each spin system. Increasing $T_{\text{e}}^{(0)}$ leads to a reversal of the itinerant spin $s_\|$ and thus a TFS on the timescale of the exchange scattering. 
This behavior occurs because the exchange scattering redistributes the deposited energy as well as angular momentum between the itinerant and localized spins all the while satisfying both conservation of total energy and total angular momentum. In Fig.~\ref{fig:Exchange_only}(b) the Stoner parameter is increased to $U=500$\,meV, which is indicative of  a more rigid itinerant carrier magnetism. Correspondingly, for the smaller initial temperatures $T_{\text{e}}^{(0)}$ the demagnetization of each spin system due to exchange scattering is reduced compared to Fig.~\ref{fig:Exchange_only}(a). For the highest excitation, i.e., $T_{\text{e}}^{(0)}=2500$\,K not only a TFS occurs, but the average itinerant spin~$s_\|$ becomes larger than the average localized spin~$S_\|$. Thus the redistribution of the deposited energy during the ultrafast exchange scattering  obviously is decisive for the following behavior. Note that the exchange scattering alone, i.e., without spin-flip scattering, yields the initial ultrafast demagnetization and the subsequent TFS. This seems to be a more realistic scenario than that described in Ref.~\onlinecite{Mentink.2012} where spin-orbit (``relativistic'') relaxation dominates the sub-picosecond dynamics, and the exchange interaction acts on a picosecond timescale.

\begin{figure}[t]
	\centering
	\includegraphics[clip=true,trim=0.5cm 0 0.2cm 0.2cm, width=0.7\textwidth]{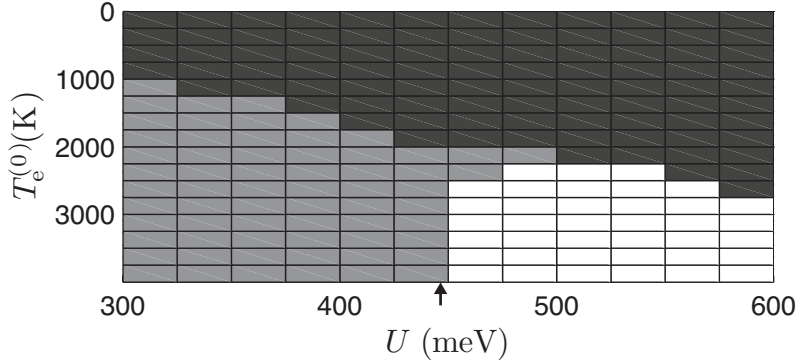}
	\caption{Color-coded dynamics: demagnetization (black), transient ferromagnetic-like state (gray) and switching (white) as a function of Stoner parameter $U$ and excitation temperature $T_{\text{e}}^{(0)}$. The exchange coupling constant is $J=100$\,meV. The arrow marks the ``critical value'' of $U=445$\,K, above which a compensation point in the equilibrium $M(T)$ relation occurs.}
	\label{fig:DTS_Tau10fs}
\end{figure} 

In Fig.~\ref{fig:DTS_Tau10fs} we collect the results from individual calculations as shown in Fig.~\ref{fig:3TempDemag}(a) and~\ref{fig:TFS_SW} by plotting the type of dynamics: demagnetization, transient ferromagnetic-like state (TFS) and switching vs.~excitation temperature $T_{\text{e}}^{(0)}$ and the strength of the itinerant ferromagnetic coupling constant $U$ in the carrier system. We find that there is a threshold for the temperature $T_{\text{e}}^{(0)}$, below which the system only de- and remagnetizes into its initial orientation (without entering a TFS). Above this threshold it depends on the Stoner parameter $U$ whether ``only'' a TFS occurs or whether heat-induced switching is achieved. Fig.~\ref{fig:DTS_Tau10fs} clearly shows that  switching only occurs for Stoner parameters $U > (450 \pm 25) \, \mathrm{meV}$. With the help of the Stoner parameter $U$ we can, in the framework of our mean-field model, put the criterion for switching in correspondence with the temperature dependence of the equilibrium magnetization $M(T)$: We find that a compensation point occurs in the $M(T)$ relation only for Stoner parameters larger than $U \simeq 445 \, \mathrm{meV}$. Within the accuracy of our numerical study these values are identical, and one can speculate about the importance of the existence of a compensation point for heat-induced magnetic switching (assuming that one starts below the compensation temperature). The present results should be taken into account when analyzing optically induced magnetization dynamics in ferrimagnetic alloys, where both heat-induced and all-optical processes may be important.~\cite{Alebrand.2012,Mangin.2014}

\section{Conclusion}

We presented a dynamical calculation of exchange scattering after spin-conserving instantaneous heating (designed to model ultrafast optical excitation) in a simple quantum-mechanical mean-field model of a ferrimagnetic alloy. We found that the exchange scattering provides the essential contribution to the ultrafast switching dynamics and established the importance of the Stoner parameter $U$ for the occurrence of a transient-ferromagnetic state and/or switching. We speculated that the occurrence of ultrafast magnetization switching may be related to the existence of a compensation point in the equilibrium magnetization.

\bibliographystyle{apsrev4-1}
\bibliography{SwitchingBib}

\end{document}